%Paper: hep-th/9411165
%From: Mark Evans <evans@prufrock.Rockefeller.EDU>
%Date: Tue, 22 Nov 94 19:25:14 -0500

\overfullrule=0pt
%\magnification=\magstep1
\baselineskip=12 pt
%A SET OF MACROS FOR WRITING PAPERS.
%
%\today GIVES TODAY'S DATE.
%
\def\today{\ifcase\month\or January\or February\or March\or April\or May\or
June\or July\or August\or September\or October\or November\or December\fi
\space\number\day, \number\year}
%
%\note{footnote} GIVES SEQUENTIALLY NUMBERED FOOTNOTES.
%
\newcount\notenumber

\def\note{\global\advance\notenumber by 1 \footnote{$^{\the\notenumber}$}}
%
%\numbereq SEQUENTIALLY NUMBERS EQUATIONS ON THE RIGHT (number)
%
\newif\ifsectionnumbering
\newcount\eqnumber
\def\cleareqnumber{\eqnumber=0}
\def\numbereq{\global\advance\eqnumber by 1
\ifsectionnumbering\eqno(\the\secnumber.\the\eqnumber)
\else\eqno(\the\eqnumber) \fi}
\def\eqalinno{{\global\advance\eqnumber by 1}
\ifsectionnumbering(\the\secnumber.\the\eqnumber)
\else(\the\eqnumber)\fi}
\def\name#1{\ifsectionnumbering\xdef#1{\the\secnumber.\the\eqnumber}
\else\xdef#1{\the\eqnumber}\fi}
\def\nosectionnumbering{\sectionnumberingfalse}
\sectionnumberingtrue
%
%\ref{\name} GIVES SEQUENTIALLY NUMBERED REFERENCES [number],
%AND ASSIGNS
%THAT NUMBER TO A MACRO \name AND WRITES REF. TO FILE 1.
%
\newcount\refnumber

\immediate\openout1=refs.tex
\immediate\write1{\noexpand\frenchspacing}
\immediate\write1{\parskip=0pt}
\def\ref#1#2{\global\advance\refnumber by 1%
[\the\refnumber]\xdef#1{\the\refnumber}%
\immediate\write1{\noexpand\item{[#1]}#2}}
\def\tie{\noexpand~}

%
% NEW SECTION: \newsection The Method. (terminate with a .)
%
\font\twelvebf=cmbx10 scaled \magstep1
\newcount\secnumber

\def\newsection#1.{\ifsectionnumbering\cleareqnumber\else\fi%
	\global\advance\secnumber by 1%
	\bigbreak\bigskip\par%
	\line{\twelvebf \the\secnumber. #1.\hfil}\nobreak\medskip\par}
%
%
%\Box GIVES WAVE OPERATOR, OR LAPLACIAN
%
\def \sqr#1#2{{\vcenter{\vbox{\hrule height.#2pt
	\hbox{\vrule width.#2pt height#1pt \kern#1pt
		\vrule width.#2pt}
		\hrule height.#2pt}}}}

%
%
%\twocolumns GIVES TWO-COLUMN OUTPUT
%
\newdimen\fullhsize
\def\fiddle{\fullhsize=6.5truein \hsize=3.2truein}
\def\fullline{\hbox to\fullhsize}
\def\mkhdline{\vbox to 0pt{\vskip-22.5pt
	\fullline{\vbox to8.5pt{}\the\headline}\vss}\nointerlineskip}
\def\mkftline{\baselineskip=24pt\fullline{\the\footline}}
\let\lr=L \newbox\leftcolumn
\def\twocolumns{\fiddle
	\output={\if L\lr \global\setbox\leftcolumn=\columnbox
		\global\let\lr=R \else \doubleformat \global\let\lr=L\fi
		\ifnum\outputpenalty>-20000 \else\dosupereject\fi}}
\def\doubleformat{\shipout\vbox{\mkhdline
		\fullline{\box\leftcolumn\hfil\columnbox}
		\mkftline} \advancepageno}
\def\columnbox{\leftline{\pagebody}}
\voffset = 17.8 mm
\hoffset = 21.6 mm
\vsize= 193 mm
\hsize= 122 mm
\parindent = 3 mm
\def\pr#1 {Phys. Rev. {\bf D#1\tie }}
\def\pl#1 {Phys. Lett. {\bf #1B\tie }}
\def\prl#1 {Phys. Rev. Lett. {\bf #1\tie }}
\def\np#1 {Nucl. Phys. {\bf B#1\tie }}
\def\ap#1 {Ann. Phys. (NY) {\bf #1\tie }}
\def\cmp#1 {Commun. Math. Phys. {\bf #1\tie }}
\def\imp#1 {Int. Jour. Mod. Phys. {\bf A#1\tie }}
\def\tie{\noexpand~}
\def\ov{\overline}
\def\s{(\sigma)}

\def\sp{(\sigma ')}

\def\Tb{\overline T}
\def\d{\delta(\sigma-\sigma ')}
\def\dpr{\delta'(\sigma-\sigma ')}

\def\dpppr{\delta'''(\sigma-\sigma ')}

\def\ints{\int d\sigma\,}

\def\dx{\partial X}
\def\dbx{{\overline\partial}X}
\def\dsx{\partial^2X}
\def\dbsx{{\overline\partial}^2X}
\parskip=0pt plus 2pt minus 0pt
%\headline{\ifnum \pageno>1\it\hfil  An
%Infinite-Dimensional Symmetry
%	$\ldots$\else \hfil\fi}
\font\title=cmbx10 scaled\magstep1
\font\tit=cmti10 scaled\magstep1
\font\eightroman=cmr8
%\footline{\ifnum \pageno>1 \hfil \folio \hfil \else
%\hfil\fi}
%\nopagenumbers
\nosectionnumbering
\raggedbottom
\font\title=cmbx10 scaled\magstep1
\font\tit=cmti10 scaled\magstep1
\footline{\ifnum \pageno>1 \hfil \folio \hfil \else
\hfil\fi}
\rightline{\vbox{\hbox{RU-94-B-8}
\hbox{CTP-TAMU-64/94}\hbox{ACT-22/94}\hbox{hep-th/9411165}}}
\vfill
\centerline{\title SYMMETRY PRINCIPLES FOR}
\vskip 20pt
\centerline{\title STRING THEORY}
\vfill
\centerline{\title Mark Evans\footnote{$^{\dag}$}
{\tenrm Work supported in part
by DOE Contract Number DOE-ACO2-87ER40325, TASK~B.}${}^{(a)}$
and Ioannis Giannakis\footnote*{\tenrm Work
supported in part by DOE Contract Number
DE-FG05-91-ER-40633.}$^{{}(b),(c)}$}
\medskip
\centerline{$^{(a)}${\tit Physics Department, Rockefeller
University}}
\centerline{\tit 1230 York Avenue, New York, NY
10021-6399}
\medskip
\centerline{$^{(b)}${\tit Center for Theoretical Physics,
Texas A{\&}M
University}}
\centerline{\tit College Station, TX 77843-4242}
\medskip
\centerline{$^{(c)}${\tit Astroparticle Physics Group,
Houston Advanced
Research Center (HARC)}}
\centerline{\tit The Mitchell Campus, Woodlands, TX 77381}
\vfill
\centerline{\title Abstract}
\bigskip
{\narrower\narrower
The gauge symmetries that underlie string theory arise from inner
automorphisms of the algebra of observables of the associated
conformal field theory. In this way it is possible to study broken
and unbroken symmetries on the same footing, and exhibit an
infinite-dimensional supersymmetry algebra that includes space-time
diffeomorphisms and an infinite number of spontaneously broken
level-mixing symmetries. We review progress in this area, culminating
in the identification of a weighted tensor algebra as a subalgebra
of the full symmetry. We also briefly describe outstanding problems.
Talk presented at the Gursey memorial conference, Istanbul, Turkey,
June, 1994.
\par}
\vfill\vfill\break

\noindent
{\title Symmetry Principles for String Theory}

\vskip 1 cm \noindent
Mark Evans${}^1$ and Ioannis Giannakis${}^{2,3}$

\bigskip\noindent
{\eightroman ${}^1$ Department of Physics, The Rockefeller
University, 1230 York Av., New York, NY 10021.\hfill\break
${}^2$  Center for Theoretical Physics, Texas A{\&}M
University, College-Station, TX 77843-4242.\hfill\break
${}^3$  Houston Advanced
Research Center, The Mitchell Campus, Woodlands, TX 77381}

\vskip 1cm
\noindent
In this talk we shall review work \ref{\evao}{M. Evans
and B. Ovrut \pr41 (1990), 3149;
\pl231 (1989), 80.}, \ref{\egw}{M. Evans and I. Giannakis,
\pr44 (1991), 2467.}, \ref{\egn}{M. Evans, I. Giannakis and D.
Nanopoulos, \pr50 (1994), 4022, hep-th/9401075.}
on deformations of
conformal field theories and symmetries of string theory.
For more details
the reader is referred to the original papers, or the review
contained in \ref\dgmtalk{M.\tie Evans and I.\tie Giannakis in
S.\tie Catto and A.\tie Rocha, eds., {\it Proceedings of the
XXth International Conference on
Differential Geometric Methods in Theoretical Physics}, World
Scientific, Singapore (1992), p.\tie 759, hep-th/9109055.}.

What is the nature of the problem? String theory is presented as a
set of rules for calculating scattering amplitudes rather than as
the the dynamical consequence of a symmetry principle. Nevertheless,
there are persuasive reasons to believe that string theory {\it
does\/} possess such an underlying symmetry, and it is surely
important that we understand what it is. This problem is not yet
solved, but we shall describe what we believe to be considerable
progress on this problem, culminating in the identification of a
new type of infinite-dimensional supersymmetry algebra, a {\it
weighted tensor algebra}, as, at least, a subalgebra of the gauge
symmetry of string theory.

To study symmetries, we seek transformations of the
space-time fields that take one solution of the classical
equations of motion to another that is physically equivalent.
Since, ``Solutions of the classical equations
of motion," are two-dimensional conformal
field theories,
we are thus interested in {\it isomorphic conformal
field theories}.

Any quantum mechanical theory (including a CFT) is defined
by three elements: i)~an algebra
of observables, $\cal A$ (determined by the degrees of
freedom of the theory
and their equal-time commutation relations),  ii)~a
representation of that algebra
and iii)~a distinguished element of $\cal A$---the Hamiltonian.
Note that for the same $\cal A$ we may
have many choices
of Hamiltonian, so that $\cal A$ should more properly be
associated with a
{\it deformation class\/} of theories than with one
particular theory.
For a CFT, we further want $\cal A$ to be generated by
{\it local\/} fields,
$\Phi(\sigma)$ (operator valued distributions on a circle
parameterized
by $\sigma$), and we require not just a single
distinguished operator, but
two distinguished {\it fields}, $T(\sigma)$ and $\ov T(\sigma)$,
which are the components of the stress tensor.
They generate the conformal transformations, and so must satisfy
the Virasoro$\times$Virasoro algebra:
$$
\eqalignno{[T\s, T\sp]&={- i c \over 24\pi}\dpppr
+2 i T\sp\dpr -  i T'\sp\d&{\global\advance\eqnumber by 1}
(\the\eqnumber a)\name{\eqvir}\cr
[\Tb\s,\Tb\sp]&={ i c \over 24\pi}\dpppr
-2 i\Tb\sp\dpr +  i\Tb'\sp\d&(\the\eqnumber
b)\cr
[T\s,\Tb\sp]&=0.&(\the\eqnumber c)\cr}
$$
Also of interest are the
{\it primary fields\/} of dimension $(d,\ov d)$, $\Phi\s \in \cal A$,
defined by the conditions
$$
\eqalign{[T\s,\Phi_{(d,\overline d)}\sp]&=i d
\Phi_{(d,\overline d)}\sp
\dpr-( i/\sqrt2)\partial\Phi_{(d,\overline d)}\sp\d\cr
[\Tb\s,\Phi_{(d,\overline d)}\sp]&=- i\overline d
\Phi_{(d,\overline d)}
\sp\dpr-( i/\sqrt2)\overline\partial\Phi_{(d,\overline
d)}\sp\d\cr}
\numbereq\name{\eqprim}
$$

Clearly, then, two CFTs will be physically equivalent if
there is an isomorphism
between the corresponding algebras of observables, $\cal
A$, that maps
stress tensor to stress tensor. (The mapping of primary to
primary is then
automatic).
The simplest such example
is a similarity transformation:
$$
\Phi(\sigma)\mapsto
e^{ih}\Phi(\sigma)e^{-ih}\numbereq\name{\eqinaut}
$$
for any fixed operator $h$. {\it Thus the physics will be
unchanged if we change
a CFT's stress tensor by just such a similarity
transformation.}

Now, the stress tensor is parameterized by the space-time
fields of the
string. For example,
$$
T_{G_{\mu\nu}}(\sigma) = \textstyle {1\over2} G_{\mu\nu}(X)
\partial X^\mu\partial X^\nu
\numbereq\name{\eqgravback}
$$
corresponds to the space-time metric $G_{\mu\nu}$, with
other fields vanishing.
If a similarity transformation (\eqinaut)
applied to $T$
generates a shift
in the space-time fields, that shift
will be a {\it symmetry\/}
transformation.

So how do $T$ and $\bar T$ deform as
we change the background fields? (we
now consider
deformations which, while they preserve conformal
invariance,
{\it need not be symmetries, e.g.} we may deform flat empty space
so that a weak gravitational wave propagates through it).
It is straightforward to
show that, to first order, the Virasoro algebras (\eqvir)
are preserved by
so-called {\it
canonical
deformations} [\evao] (see also \ref{\findef}{G. Pelts,
Rockefeller University preprints RU-93-5-B, hep-th/9309141 and
RU-94-5-B, hep-th/9406022 and contribution, this volume.}),
$$
\delta T\s = \delta\Tb\s =
\Phi_{(1,1)}\s \numbereq\name{\eqcan}
$$
where $\Phi_{(1,1)}\s$ is a primary
field of dimension (1,1). We reiterate:
(\eqcan) does {\it not\/} in general correspond to a
symmetry transformation,
but it preserves conformal invariance.
Since (1,1) primary
fields are vertex operators for physical states, they correspond
naturally to the space-time fields, and equation
(\eqcan) makes transparent the
connection between changes of the stress tensor and
changes of the space-time
fields. Returning now to the problem of
symmetries, if we take
the generator $h$ in equation (\eqinaut) to be the zero
mode of an
infinitesimal (1,0) or (0,1) primary field (a current),
then it is
a simple consequence  of the definitions
that its action on the stress tensor is necessarily
a canonical deformation,
and so may be translated directly into a change in the
space-time fields (for examples, see [\evao]). It is
well known that {\it conserved\/} currents generate
symmetries \ref{\band}{T. Banks and L. Dixon
\np307 (1988), 93.}, but
within the formalism described here, conservation is {\it
not\/}
necessary, a fact that does not seem to have been widely
appreciated.
Indeed, it may be shown that a non-conserved current
generates
a symmetry that is spontaneously broken by the particular
background being
considered [\evao].

At this point it is convenient to assess the strengths and weaknesses
of the analysis so far. We begin with the strengths:{ \item{$\bullet$}
The fact that a current zero-mode generates a canonical deformation
guarantees that we can translate the inner automorphism into a
transformation on the physical space-time fields.  \item{$\bullet$}
In this way, we may exhibit symmetries both familiar (general
coordinate invariance, among many \ref{\ovre}{M.\tie Evans and
B.\tie Ovrut, \pl214 (1988), 177; \pr39 (1989), 3016; M.\tie Evans,
J.\tie Louis and B.\tie Ovrut, \pr35 (1987), 3045.}) and unfamiliar
(an infinite class of spontaneously broken, level-mixing gauge
supersymmetries) [\evao].\par} \smallskip \noindent However, there
are also deficiencies:  {\item{$\bullet$} Explicit calculations are
hard to perform for a general configuration of the space-time fields.
It would appear that we need to know the precise form of the
stress-tensor and its currents for that general configuration (recall
that a current is only a current relative to a particular stress-tensor).
This is a very hard problem, essentially requiring us to find the
general solution to the equations of motion before we can discuss
symmetries.  \item{$\bullet$} It is hard to say anything about the
symmetry algebra.  This is just the algebra of the generators, but
it is {\it not\/} in general true that the commutator of two current
zero-modes is itself a current zero-mode: our symmetry ``algebra"
does not close!  \item{$\bullet$} Finally, by considering a few
examples, it is easy to see that the canonical deformation of
equation (\eqcan) corresponds to turning on space-time fields {\it
in a particular gauge\/} (something like Landau or harmonic gauge).
We would like to understand the gauge principle behind string theory
without imposing gauge conditions.\par}

It turns out that these three drawbacks are intimately related, and
may be largely overcome by moving beyond canonical deformations,
equation (\eqcan), which are {\it not\/} the
most general infinitesimal deformation that preserve the
Virasoro
algebras (\eqvir). In [\egw] we showed that, for the
massless degrees
of freedom of the bosonic string in flat space, we could
find a distinct
deformation of the stress tensor for each solution of the
linearized
Brans-Dicke equations. This correspondence was found by
considering the
general translation invariant {\it ansatz\/} of naive
dimension two for
$\delta T$;
$$
\eqalignno{
\delta T=&H^{\nu\lambda}(X)\dx_\nu\dbx_\lambda+
A^{\nu\lambda}(X)\dx_\nu\dx_\lambda+\cr
&\qquad\qquad B^{\nu\lambda}(X)\dbx_\nu\dbx_\lambda
+C^\nu(X)\dsx_\nu+D^\lambda(X)\dbsx_\lambda,&
\eqalinno\name{\eqansatz}\cr
}
$$
with a similar, totally independent {\it ansatz\/} for
$\delta\Tb$.
The fields
$H^{\mu\nu}$ {\it etc.}~turn out to be characterized in
terms of solutions
to the linearized Brans-Dicke equation when we demand that
the deformation
preserves (to first order) the Virasoro algebras (\eqvir).

By considering
this more general {\it ansatz}, we get more than just
covariant equations of
motion---we also understand a larger set of symmetry
generators, $h$.
Indeed, any generator that preserves the form of the {\it
ansatz\/}
(\eqansatz)
will do.

The condition that $\delta T$ be of naive dimension two
(with which we
shall soon dispense) is preserved if $h$ is of naive
dimension zero. The
condition of translation invariance is
preserved whenever $h$ is a zero-mode.

The lesson to be drawn from this massless example is
clear: the way to
introduce space-time fields unconstrained by gauge
conditions is to consider
an {\it arbitrary translation invariant ansatz\/} for the
deformation
of the stress tensor, and to ask only that it preserve the
Virasoro
algebras. To move beyond the massless level, we simply
drop the requirement
that the naive dimension be two. Thus symmetries are
generated by arbitrary zero-modes in $\cal A$.
We argued in [\egw] that, as
with the superfield
formulation of supersymmetric field theories,
this was likely to
introduce auxiliary fields beyond the massless level, but so be it.
(Indeed, this
whole formulation
of string theory is rather akin to a superspace approach,
with
$T$ and $\Tb$ as superfields and
derivatives of the world-sheet scalars playing the r\^ole
of the
odd coordinates of superspace).

This extension of the set of symmetry generators
ameliorates each of the drawbacks mentioned above.
It should also be emphasized that the symmetry algebra is
now manifestly background independent, since being a
zero-mode is a property independent of the choice of stress-tensor.

Nevertheless, difficulties remain. In particular, the operator
algebra, $\cal A$ and its subalgebras require a more careful
construction than is usually given. In [\egn] we showed that, even
in free scalar field theory, normal ordering is inadequate to define
composite operators, even vertices and currents. We demonstrated
this by showing that the commutator of two such fields may be
singular. Typically, the algebra $\cal A$ is constructed by combining
two (non-local!) chiral subalgebras into what Zuckerman has called
the (local!) ``bilateral'' algebra \ref\zuck{G. Zuckerman, this
volume.}.  However, this construction is problematic because
commutators in one chiral subalgebra will often be very local
(involving $\delta$-functions). This puts fields from the other
chiral component at the same point which, of course, introduces
singularities [\egn].

This problem is not yet fully resolved, but it is possible [\egn]
that the correct approach is to analytically continue in space-time
momenta---the same procedure used to calculate scattering amplitudes.
However, we may avoid this difficulty by finding
interesting infinite-dimensional subalgebras of the full symmetry
where it simply does not arise. The simplest such examples are
just local subalgebras of the chiral algebras which go to make up
$\cal A$. For the uncompactified bosonic string, these are
generated by operators of the form:
$$
h= \ints {\phi_{\mu\nu\cdots\rho}}{\partial^{w_1}}X^{\mu}{\partial^{w_2}}
X^{\nu} \cdots {\partial^{w_n}}X^{\rho}. \numbereq\name\eqkorad
$$
 It is clear that each such generator is specified by a constant
 tensor, $\phi$, with a positive-definite integer weight, $w$
 associated to each index. For this reason, we coined the name {\it
 weighted tensor algebra} for such structures. It is straightforward,
 if a little tedious, to calculate the commutators of such operators.
 If we label each generator by its tensor and weights,
 $\left\{\phi_{\mu_{\cal S}}, w_{\cal S}\right\}$, where ${\cal S}$
 is some index set, then the commutator is
$$
\eqalignno{
&\bigl[\bigl\{\psi_{\mu_{\cal S}},w_{\cal S}\bigr\},
\bigl\{\chi_{\nu_{\cal T}},v_{\cal T}\bigr\}\bigr] =\cr
&\sum_{\scriptstyle\cal U\subseteq S\atop\scriptstyle
\mathstrut P\colon
U\hookrightarrow T} {\cal C_{U,P}} \Delta_{\left|\cal
S-U\right|}^{\left(-1+
\sum_{i\in\cal U} w_i+v_{{\cal P}(i)}\right)}
\left\{\psi_{\mu_{\cal S}}\chi_{\nu_{\cal T}}
\prod_{i\in\cal U} \eta^{\mu_i
\nu_{{\cal P}(i)}},\; w_{\cal S-U}\oplus v_{\cal
T-P\left(U\right)}\right\}.\qquad
&\eqalinno\name{\eqcommutator}\cr
}
$$
The sum, over subsets $\cal U$ of $\cal S$ and injections $\cal P$,
is just a formal way of summing over all possible contractions of
the tensors. The operator $\Delta$ is defined by
$$
\Delta_r \left\{\phi^{(k)}, w_i\right\} =
	\sum_{l=1}^r \left\{\phi^{(k)},
(w_i+\delta_{il})\right\},
		\numbereq\name{\eqdefdelta}
$$
and differentiates the first $r$ factors in the integrand of the
generator, (\eqkorad).  The coefficients are given by
$$
{\cal C_{U,P}} = {\prod_{k\in\cal
U}(-1)^{w_k}(w_k+v_{{\cal P}(k)}-1)!\over
	\left(\left(\sum_{k\in\cal U} w_k+v_{{\cal
P}(k)}\right)-1\right)!}
\numbereq\name{\eqstructconst}
$$

It may not be pretty, but it's a gauge subalgebra of string theory,
and it seems likely that the full algebra will have a similar form.
Somehow, we shall have to learn to love it.

\bigskip The work of M.~E. was supported in part  by DOE Contract
Number DOE-ACO2-87ER40325, TASK~B, and that of I.~G. by DOE Contract
Number DE-FG05-91-ER-40633.

\immediate\closeout1
\bigbreak\bigskip

\line{\twelvebf References. \hfil}
\nobreak\medskip\vskip\parskip

\input refs

\vfil\end